\documentclass[aps,prl,twocolumn,floatfix,showpacs,amsmath,amssymb]{revtex4}
\usepackage{graphicx}
\usepackage{amsmath}
\usepackage{amsbsy}
\usepackage{dcolumn}
\usepackage{bm}

\begin{document}

\title{
Spin Correlations and Finite-Size Effects in the One-dimensional 
Kondo Box}

\date{\today}
\author{Thomas Hand, Johann Kroha and Hartmut Monien}
\affiliation{Physikalisches Institut, Universit\"at Bonn,
Nussallee 12, 53115 Bonn, Germany}

\begin{abstract}
We analyze the Kondo effect of a magnetic impurity attached to an 
ultrasmall metallic wire 
using the density matrix renormalization group. The spatial spin
correlation function and the impurity 
spectral density are computed for system sizes of up to $L=511$ sites, 
covering the crossover 
from $L<\ell _K$ to $L > \ell _K$, with $\ell _K$ the spin screening length.
We establish a proportionality between the weight of the
Kondo resonance and $\ell_K$ as function of $L$. This 
suggests a spectroscopic way of detecting the Kondo cloud.
\end{abstract}
\bigskip
\noindent 
\maketitle 

Scanning tunneling techniques have recently allowed to
observe the Kondo effect of a magnetic atom in an ultrasmall metallic
box \cite{odom00}, possibly providing a direct probe of the 
long sought-after Kondo screening cloud. 
The Kondo effect is characterized by a narrow resonance of width $\sim T_K$, 
the Kondo temperature, at the Fermi energy $\varepsilon_F$ \cite{hewson93}. 
It is intimately related to 
the formation of a many-body singlet state, comprised of the 
impurity spin and a cloud of surrounding, spin-correlated electrons, 
the so-called Kondo spin screening cloud.
Its spatial extent is vital for the coupling between neighboring 
Kondo impurities in a metal and, hence, is at the heart of spatial magnetic 
correlations and ordering transitions in Kondo and Anderson lattices  
and also in Hubbard or $t-J$ systems, which exhibit local Kondo 
physics, as has been 
demonstrated by the dynamical mean field theory (DMFT) 
treatment of the problem \cite{georges96}.
However, while the spectral and thermodynamic features of
Kondo impurities have been well understood 
\cite{hewson93}, the structure of the Kondo cloud has remained 
controversial for a long time.
Researchers have been looking intensively for ways of
observing the Kondo cloud. These include the Knight shift 
\cite{slichter74} and recently theoretical investigations of
the persistent current \cite{affleck01} or the conductance
\cite{affleck02} in mesoscopic Kondo systems. For about 25 years 
it was generally believed, and in the 1990s supported by
scaling arguments \cite{affleck96}, that the Kondo cloud is
characterized by a single length scale, $\xi_K=\hbar v_F/T_K$. 
It is the spin coherence length, i.e. 
the distance traveled by a scattered electron
with Fermi velocity $v_F$, until the impurity spin 
(whose lifetime is $\hbar/T_K$) flips.
Although $\xi _K$ can reach almost macroscopic values  
($\xi_K \approx 10^3 k_F^{-1}$ for $T_K=1\ K$, $k_F$ being the Fermi 
wave number), it has never been observed in experiments.\newline
Only recently it was realized that another length scale, $\ell_K$, 
arises in a $d$-dimensional Kondo system, if all conduction electron 
states couple equally to the impurity spin \cite{thimm99}. 
It is the length of a
finite-size conduction electron sea, the ``Kondo box'', which is
so small that its level spacing $\Delta$ is comparable to $T_K$ of the 
bulk system and cuts off the logarithmic Kondo correlations.
Therefore, a box of length $\ell_K$ sustains just 
one conduction electron state within the Kondo scale $T_K$ to form the 
Kondo singlet \cite{affleck00}, i.e. $\ell_K$ is the size of the Kondo
cloud, the Kondo screening length.
Equating $\Delta = T_K$, with $\Delta$ the inverse of the
typical density of states (DOS) in a box of size $\ell _K$,
$N (\ell_K) = (\ell _K/2\pi)^d S_d k_F^{d-1}/(\hbar v_F)$,
yields, 
\begin{equation}
\ell _K = 2\pi \left( \xi_K/S_d k_F^{d-1}\right)^{1/d}\ ,
\label{eq:screeninglength}
\end{equation}
with $S_d$ the surface of the 
$d$-dimensional unit sphere \cite{thimm99}. 
Hence, 
$\ell _K$ is an intermediate length scale, which for $d\geq 2$ can be
substantially smaller than the
coherence length, $1/k_F <\ell_K<\xi_K$, and 
$\ell_K \approx \xi_K$ only in effectively 1D systems.
Another length scale, $\ell_{RKKY}$, would arise in dilute 
Kondo systems as the one when the RKKY coupling between neighboring
impurities equals $T_K$, 
$\ell_{RKKY}=[JN(k_F^{-1})]^{1/d}\ell_K<\ell_K$ \cite{affleck00},
where $JN(k_F^{-1})$ is the dimensionless spin coupling. 
The different physical meaning of $\xi_K$ and $\ell_K$ should be kept
in mind for the design of related experiments. For example, 
experiments to detect the Kondo
cloud via finite system size, like those proposed in 
Refs.\ \cite{affleck01,affleck02}, probe $\ell_K$ rather than $\xi_K$. 
These experiments should be performed on 1D wires in order for
$\ell _K$ to be in an experimentally accessible range.
1D Kondo boxes have up to now been realized as ultrashort
Carbon nanotubes \cite{odom00}, which, however, do not easily 
permit persistent \cite{affleck01} or 
transport \cite{affleck02} current measurements.

In this work we show numerical evidence that the Kondo cloud 
can be detected via 
spectroscopy of the Kondo resonance in a 1D Kondo box. To that end
we establish a non-trivial proportionality between the Kondo spectral
weight and the spin screening length as function of system size,
using large-scale density matrix renormalization group (DMRG) calculations
\cite{white92,kuehner99}.
The systems considered here are 1D in the sense that the magnetic
impurity is side-coupled to a finite chain of atoms 
only at a {\it single} site $x_0$ of the chain. This is different from the
ultrasmall boxes considered in Refs.\ \cite{thimm99,schlottmann02}, 
where the effective hybridization was the same for all 
states in the box. The latter systems may have
been realized most recently in molecules
\cite{booth05}. As a result of the local coupling we observe strong
mesoscopic variations of $T_K$ and of the spectral features. 
We analyze, under which mesoscopic conditions the above-mentioned 
proportionality prevails. 

\begin{figure}[t]
\begin{center}
\includegraphics[width=0.9\linewidth]{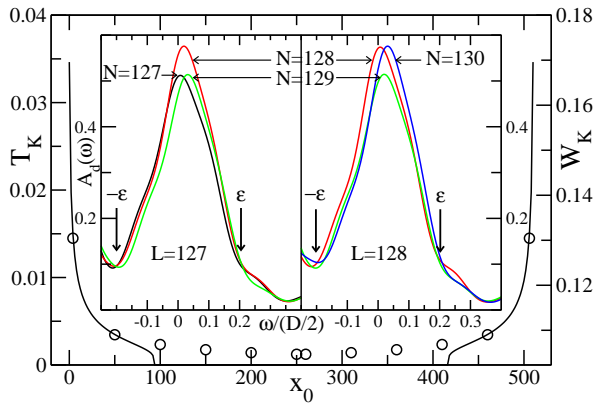}
\end{center}
\vspace*{-0.4cm}
\caption{\label{fig1} (Color online)
Finite-size and even/odd effects in the 1D Kondo box.
Solid line: $T_K$ as a function of impurity position $x_0$
for  $L=511$, $N=512$, $V=0.35$ and open boundary conditions,
as obtained from Eq.\ (\ref{eq:TK}). Open circles: 
Weight of the Kondo peak, $W_K$, as defined in the text and in the
inset. The results shown are for even $x_0$ only (see text). 
The inset shows the Kondo peak for $V=0.35$, $x_0=4$ and for
various successive values of $L$ and $N$.
}
\vspace*{-0.3cm}
\end{figure}

The Hamiltonian for an Anderson impurity with local energy 
$\varepsilon _d$ and on-site Coulomb
repulsion $U$, side-coupled via the hybridization matrix element $V$
to the site $x_0$ on a 1D chain of $L$ sites, reads,
\begin{eqnarray}
H&=& H_{ch}
 + \varepsilon_d 
  \sum _{\sigma} d_{\sigma}^{\dagger}d_{\sigma}^{\phantom{\dagger}}
 + V \sum _{\sigma} 
     \left[ c_{x_{o}\sigma}^{\dagger} d_{\sigma}^{\phantom{\dagger}} 
     + H.c.\right] 
\nonumber\\
 &+& U d_{\uparrow}^{\dagger} d_{\uparrow}^{\phantom{\dagger}}
     d_{\downarrow}^{\dagger}d_{\downarrow}^{\phantom{\dagger}}\ ,
\label{eq:hamiltonian}
\end{eqnarray}
where $H_{ch}= -t \sum _{\langle i,j\rangle ,\sigma} 
       c_{i\sigma}^{\dagger} c_{j\sigma}^{\phantom{\dagger}}$,
$i,j=0,\dots ,L-1$, is the free chain Hamiltonian
with nearest-neighbor hopping $t>0$. For the evaluations 
we choose the total electron number $N$ near half bandfilling 
($N=L\pm 2$, $\varepsilon_F\approx0$) 
and use generic parameters for the model in the Kondo regime,
$\varepsilon _d=-0.55$, $U=5$, and $V$ as indicated where appropriate.
All energies are given in units of the
half bandwidth $D=2t$. The Kondo spin coupling is 
given by $J=V^2[1/|\varepsilon _d| + 1/(\varepsilon_d +U)]$.

{\it $T_K$ in finite systems.}
As mentioned above, for this realistic model one expects large finite-size 
effects, because the effective impurity-chain coupling, which 
governs the low-energy Kondo physics, depends on the 
amplitudes of the free-electron eigenfunctions of 
the chain, $\Psi_k(x_0)$, at the position $x_0$. The Kondo scale 
$T_K$ is defined as the temperature $T$ at which the 2nd order
contribution to the spin scattering T-matrix equals the 1st order  
\cite{hewson93}, a condition which in the finite system reads,  
\begin{equation}
-2 J \sum _k \frac{|\Psi_k(x_0)|^2}{\varepsilon _k -\varepsilon_F}\
     \frac{1}{e^{-(\varepsilon_k-\varepsilon_F)/T_K}+1} =1\ , 
\label{eq:TK}
\end{equation}     
with $\varepsilon _k$ the levels of the
free chain. It is seen that $T_K$ itself depends on the impurity 
position $x_0$ \cite{affleck03,morr06} and on the system size $L$ as well. 
The strong $x_0$ dependence of $T_K(x_0)$ shown in 
Fig.\ \ref{fig1} is due to the increase of the 1D local DOS towards 
the ends of a chain with open boundary conditions \cite{zarand96}. 
If $x_0$ ist too close to the center of the chain 
(e.g. $|x_0/L-1/2|\lesssim 160$ in Fig.\ \ref{fig1}), 
the log contributions in Eq.\ (\ref{eq:TK})
are cut off by the level spacing of the finite system {\it before} the
breakdown of perturbation theory, so that the system stays in the
perturbative regime for all temperatures, i.e. $T_K=0$ (Fig.~\ref{fig1}).
Hence, in an ultrasmall system the expressions for $\ell_K$ and $\xi_K$ 
discussed above can, at best, serve to obtain typical values for these
quantities. We find that the width of the Kondo resonance for various 
$\varepsilon_d$, $U$, $V$ resembles roughly $T_K$ of Eq. (\ref{eq:TK}), 
however obscured by the discreteness of the box spectrum. 
Detecting the Kondo cloud by varying the system
size then becomes a nontrivial task, since $\ell _K$ itself depends on $L$.
Detailed numerical calculations are, therefore, needed in order to
incorporate these finite size effects and to extract the universal 
features that persist under these conditions.

\begin{figure}[t]
\begin{center}
\includegraphics[width=0.85\linewidth]{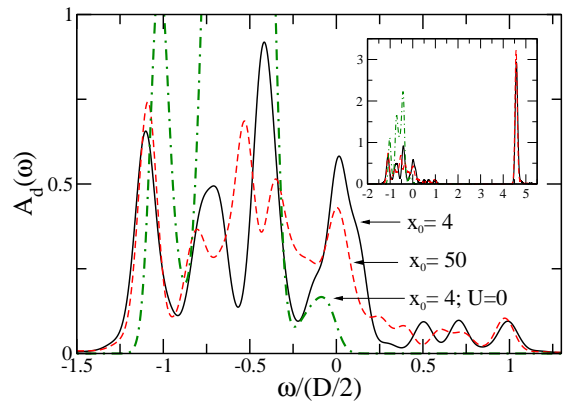}
\end{center}
\vspace*{-0.4cm}
\caption{\label{fig2} (Color online)
The impurity spectral density $A_d(\omega)$ for 
$L=127$, $N=128$, $V=0.35$ and two different $x_0$; 
$\eta =0.05$. Comparison with the non-interacting spectrum ($U=0$) 
exhibits the Kondo enhancement of the peak
near $\varepsilon _F=0$.
The inset shows the 
upper Hubbard peak near $\omega=\varepsilon_d +U$.
}
\vspace*{-0.3cm}
\end{figure}

{\it Numerical method and testing.}
Applying an efficient DMRG code \cite{hand06} 
to the model Eq.\ (\ref{eq:hamiltonian}),
we have computed the (retarded) impurity Green's function and the
equal-time spin correlation function at $T=0$,
\begin{eqnarray}
G_{d\sigma}(\omega) \hspace*{-0.2cm} &=& \hspace*{-0.2cm}
\langle 0 |
\left[ d_{\sigma}^{\phantom{\dagger}}\frac{1}{E +i\eta - H}
       d_{\sigma}^{\dagger} +
       d_{\sigma}^{\dagger}\frac{1}{E +i\eta - H}
       d_{\sigma}^{\phantom{\dagger}} 
\right]
| 0 \rangle
\nonumber\\
\label{eq:Gd}\\
K(r) &=& 
\langle 0 | S^z_{i} S^z_{x} | 0 \rangle -
\langle 0 | S^z_{i} | 0 \rangle
\langle 0 | S^z_{x} | 0 \rangle \ ,
\label{eq:spincorr}
\end{eqnarray}
respectively, for system sizes of up to $L=511$.
Here $\omega$ is the single-particle excitation 
energy relative to the many-body ground state energy $E_0$, $E=\omega+E_0$. 
$|0\rangle$ the DMRG many-body ground state
and $S^z_{i}$, $S^z_{x}$ the z-components of the 
spin-1/2 operators on the impurity and on chain site $x$, $r=x-x_0$, 
respectively. The impurity spectral density is
$A_{d\sigma}(\omega)=-\frac{1}{\pi}{\rm Im}G_{d\sigma}(\omega)$. 
Open boundary conditions are applied to facilitate convergence of the
DMRG algorithm. They also appear appropriate for a wire (weakly) coupled
to leads. For the dynamical quantities we have used
both the correction vector (CV) method \cite{kuehner99}, and the 
Lanczos method (LM) \cite{hallberg95}.
For the CV method, $m=200$ basis states were retained
in each DMRG iteration, which proved sufficient to compute the 
residue of the CV 
$(\omega +i\eta - H )^{-1}d_{\sigma}^{\dagger}|0\rangle$
with a precision of $10^{-8}$ for each $\omega$. 
For the LM we used 3 to 5 target states, kept
($m=2600$) basis states and carried 
out 200 Lanczos steps to build the Krylov subspace. The
comparison of the two methods for $L$ up to 128 yields 
excellent agreement (better than 0.1 per cent) 
for $\omega \lesssim T_K$ and still good agreement 
(better than 10 per cent) even for the highest $|\omega|\approx D$, where
the LM becomes inaccurate. Scaling up the system size
from $L=128$ to $L'=511$ reduces the frequency range where Lanczos is
accurate by a factor $L/L'$, which was satisfactory for the calculations
in the Kondo regime. For the largest systems ($L=511$) 
we, therefore, used the numerically less demanding LM.

\begin{figure}[t]
\begin{center}
\includegraphics[width=0.9\linewidth]{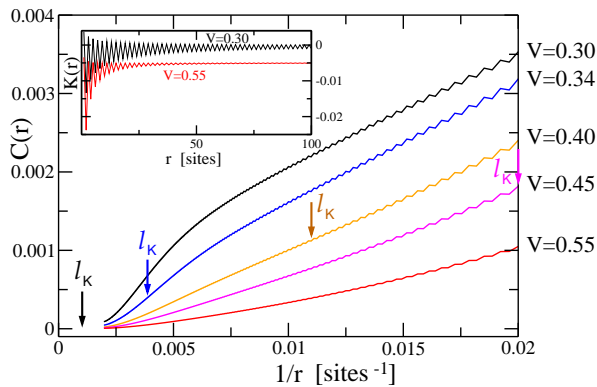}
\end{center}
\vspace*{-0.4cm}
\caption{\label{fig3} (Color online)
The average $C(r)$ of the spin correlation function $K(r)$
is shown as function of $1/r$ for $L=511$,
$N=512$ and various hybridization strengths $V$;
$r=x-x_0$, $x_0=4$. 
Inset: $K(r)$ showing RKKY oscillations. The $V=0.55$ curve is
offset by -0.005 for clarity.
}
\vspace*{-0.3cm}
\end{figure}

Note that all DMRG calculations are done in the canonical
ensemble with fixed electron number $N$ 
and fixed total spin $S$, whereas experimental systems are usually
coupled to a particle reservoir. 
Life-time effects of $N$ and $S$ are included as a Lorentzian
(for the CV method) or Gaussian (for the LM)
broadening $\eta$ of the energy levels, with $\eta = 0.05$ below.
$x_0$ is chosen near the end of the chain, where
$T_K$ is large enough 
(see above and Fig.\ \ref{fig1}) so that we can sweep through the
crossover from $\Delta > T_K$ to $\Delta < T_K$. Furthermore
we choose $x_0$ to be even, because on all odd sites the chain wave 
function at $\varepsilon _F=0$ has a node, so that for small $L$ 
($\Delta > T_K$) the impurity would be decoupled.

\begin{figure}[t]
\begin{center}
\includegraphics[width=0.9\linewidth]{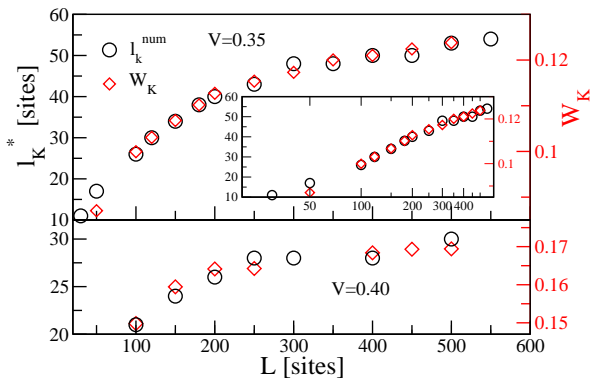}
\end{center}
\vspace*{-0.4cm}
\caption{\label{fig4} (Color online)
The weight of the Kondo resonance $W_K$ and the numerically 
determined screening length $\ell _K^{*}$ 
as  function of size $L$ for $x_0=4$ and $N=L+1$, $N$ odd. 
}
\vspace*{-0.3cm}
\end{figure}

{\it Results.}
The $T=0$ impurity spectrum $A_{d\sigma}(\omega)$ shown in Fig.\ 
\ref{fig2} exhibits a rich multiple peak structure 
even in the single-particle spectral weight near 
$\omega \approx \varepsilon_d$, induced by the discrete local
conduction electron spectrum even for the largest $L$, when the 
impurity is placed close to the boundary. The Kondo peak
is identified in Fig.\ \ref{fig2} as the one near $\omega=0$ through its 
systematically increasing weight as the interaction $U$ is switched 
on, as $L$ is increased (Fig.\ \ref{fig4}), or 
as $T_K$ is increased by moving the impurity from $x_0=50$ to $x_0=4$  
(see also Fig.\ \ref{fig1}). 
The latter would correspond to decreasing $T$ \cite{thimm99} in a 
temperature dependent measurement. For the local impurity coupling $V$ 
in Eq.\ (\ref{eq:hamiltonian}) we find that the particle number 
parity effect in the {\it position} of the spectral features
(1 or 2 peaks within $|\omega| \lesssim T_K$) is essentially washed out by
finite-size irregularities of the local conduction electron spectrum
even for small level broadening $\eta$ (not shown), 
in contrast to the pronounced even/odd characteristics predicted for 
equal coupling to all conduction states \cite{thimm99}. 
However, the even/odd effect is seen in the inset of Fig.~\ref{fig1} 
as an enhancement of the Kondo peak for even as compared to odd $N$ 
for fixed system size $L$. 

The impurity-conduction electron spin correlation function $K(r)$, 
as computed from Eq.\ (\ref{eq:spincorr}), is shown in the inset 
of Fig.\ \ref{fig3}. It displays RKKY oscillations with 
period $\lambda_{RKKY}=\pi/k_F=2a$ ($a=$lattice constant), 
Its overall weight yields $s^{-2}\sum _r K(r)= - n_{d}$,
with $n_{d}$ the total impurity occupation number,
confirming complete screening of the impurity spin, $s=1/2$.
The average $C(r)=(K(r)+K(r+1))/2$ measures the spin content in the
Kondo cloud at distance $r$, while $\Delta K(r)=|K(r)-K(r+1)|/2$
is the amplitude of the RKKY oscillations. $C(r)$ is shown in Fig.\ \ref{fig3},
together with the respective $\ell _K$ as calculated from 
Eqs.\ (\ref{eq:screeninglength}), (\ref{eq:TK}).  
The expected $1/r^{d}$ behavior \cite{affleck96}  
is clearly seen for $1/k_F \ll r<\ell_K$ and $V=0.3,\ 0.34$. 
For smaller $\ell _K$ ($V=0.4$, $0.45$, $0.55$) the powerlaw range 
is too narrow to be observable. For $r \stackrel {>}{\sim} \ell _K$, 
we find exponential decay, $C(r) \propto \exp(-2r/\ell _K)$ 
(Fig.\ \ref{fig3}), 
and similar for $\Delta K(r)$. This is expected for the correlator of two 
non-conserved quantities, like $S_i^z$, $S_x^z$, with a finite correlation 
length. In the asymptotic region, $r\gg \ell _K$, the exponential 
behavior should be overidden by the slower powerlaw decay,
$C(r) \propto 1/r^{d+1}$, expected from general Fermi liquid 
arguments \cite{ishii78,affleck98}. The numerical data show indications
of this crossover for the largest $L$ and the smallest $\ell _K$.
A more detailed analysis of the complex $r$-dependence 
will be presented elsewhere.

\begin{figure}[t]
\begin{center}
\includegraphics[width=0.85\linewidth]{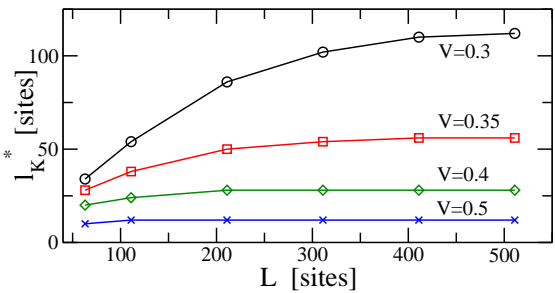}
\end{center}
\vspace*{-0.4cm}
\caption{\label{fig5} (Color online)
$W_K$ like in Fig.\ \ref{fig4},
but for $N$ even.
}
\vspace*{-0.3cm}
\end{figure}

For $V=0.3$ the $\ell_K$ from Eq.\ (\ref{eq:screeninglength}) is $936 > L$. 
$C(r)$ is then not cut off by $\ell_K$ but by $L$, and the conduction 
electron
spin density necessary for complete spin screening is accumulated at 
shorter distances, leading to a positive $y$-axis
intersection, see Fig.\ \ref{fig3}. This displays the difficulty in 
extracting $\ell _K$ directly from finite systems and the limited 
applicability of Eqs.\ (\ref{eq:screeninglength}), (\ref{eq:TK}) for this
purpose. 
Therefore, we combine the results for $A_{d\sigma}(\omega)$ and $C(r)$ to 
obtain an experimental signature of the (bulk) screening length
$\ell _K$ in the finite-size spectra.
In doing so one must observe that for $L\lesssim \ell _K$, 
$\ell _K$ itself becomes size and position dependent according to 
Eqs.\ (\ref{eq:screeninglength}), (\ref{eq:TK}) and that for our
system with fixed total spin there is always a total spin $1/2$
in the cloud, no matter how small $L$.  
Therefore, we define the screening length of the finite 
system, $\ell_K^{*}(L)$, by the volume needed
to host a certain fraction $c$ of the total spin,
$s^{-2}\int _0^{\ell_K^{*}} dr K(r) = c$, where $s=1/2$ is the 
electron spin. The Kondo spectral weight is defined as 
$W_K(L) = \int _{-\varepsilon '}^{\varepsilon} d\omega A_{d\sigma}(\omega)$
(c.f. Fig. \ref{fig1}, inset),
where the boundaries $-\varepsilon '$, $\varepsilon$, are chosen so 
as to cover the Kondo resonance, identified numerically as that
part of the spectrum around $\varepsilon _F$ which increases as $U$ is
switched on (c.f. Fig.\ \ref{fig2}).
The results for both quantities are shown in Figs.\ \ref{fig4}, \ref{fig5}
for $c=0.75$ and $\varepsilon =\varepsilon ' =0.2$. 
For odd particle number $N$ (Fig.\ \ref{fig4}) the nontrivial 
proportionality 
$\ell_K^{*}(L)/W_K(L)=\alpha (J)$ for the complete range of $L$
is established. We checked that it persists independent of the precise 
choice of $\varepsilon$, $\varepsilon '$ and $c$. 
Both $W_K(L)$ and $\ell_K^{*}(L)$ are logarithmically
suppressed with decreasing $L$ (inset of Fig.\ \ref{fig4}). 
For universality reasons we expect the proportionality to extend out
to $L\to\infty$, where $\ell_K^{*}(L)\to\ell_K(\infty)$ must saturate at its
bulk value. The proportionality $W_K(L)\propto\ell_K^{*}(L)$ 
persists for different values of $J$ (Fig.\ \ref{fig4}), and the corresponding 
$\ell _K^{*}(L)$ can be scaled on top of each other by plotting 
$\ell _K^{*}(L)/\ell _K^*(J)$ vs $L/\ell _K^*(J)$, with the scaling 
parameter $\ell_K^*(J)\approx \ell_K(\infty)$.  
The above relation can be used to determine $\ell_K^{*}(L)$
by a spectroscopic measurement and to extrapolate to its bulk value, once the 
proportionality constant $\alpha (J)$ is determined.
Fig.\ \ref{fig5} displays $\ell_K^{*}$ for even $N$, showing an earlier 
saturation compared to Fig.\ \ref{fig4}, as expected from the even/odd effect
\cite{thimm99}. However, we find $W_K(L) \approx const.$ in this case,
breaking the above proportionality. By an analysis of the spectra this is
traced back to the fact that for the parameters of Fig.\ \ref{fig5} the 
impurity spectrum is dominated by a strong $L$-independent 
{\it single-particle} peak inherited from the free conduction band, 
while the spin structure, $C(r)$, retains its $L$-dependence. 
This is to emphasize that it is essential to identify the $\varepsilon_F$ peak
as a Kondo peak first, e.g. by its logarithmic $T$ or $L$ dependence, 
before the above analysis can be applied.

To conclude,
we have analyzed the spectral and the 
spin structure of ultrasmall Kondo
systems in the presence of strong finite-size fluctuations and 
even/odd effects using DMRG. Despite these non-universal effects 
we have identified a procedure to measure the spin screening length 
$\ell _K$ by tunneling spectroscopy, e.g. on carbon nanotube Kondo boxes.
Further research is needed to understand the relation
$\ell_K^{*}(L)=\alpha W_K(L)$ and to determine the 
proportionality factor $\alpha (J)$.

We acknowledge useful discussions with I. Affleck and S. White.
This work was supported in part 
by DFG through grants KR1726/1, SFB 608, and SP1073. 

\vspace*{0.8cm}

\hspace*{-0.2cm}Electronic address: kroha@physik.uni-bonn.de

\end{document}